\documentclass{article}
\usepackage{spconf,graphicx}
\usepackage{amssymb}
\usepackage{amsmath}
\usepackage{float}
\usepackage[dvipsnames]{color, xcolor}
\usepackage{url}
\usepackage{multirow, booktabs, threeparttable}
\usepackage{stfloats}
\usepackage{wrapfig}
\usepackage{setspace}

\title{VoiceLens: Controllable Speaker Generation and Editing with Flow}
\name{
Yao Shi, Ming Li
\sthanks{
	Corresponding author: Ming Li.
	This research is funded in part by the National Natural Science Foundation of China (62171207) and Science and Technology Program of Guangzhou City (202007030011) and Bytedance. Many thanks for the computational resource provided by the Advanced Computing East China Sub-Center.
}
}

\address{
School of Computer Science, Wuhan University, Wuhan, China \\
Data Science Research Center, Duke Kunshan University, Kunshan, China \\
{\url{ming.li369@dukekunshan.edu.cn}}
}

\begin{document}
% \ninept
%
\maketitle
\begin{abstract}
%\textbf{Fake abstract.} In this paper, we present AISHELL-3 , a large-scale multispeaker Mandarin speech corpus which could be used to train multi-speaker Text-To-Speech (TTS) systems. The corpus contains roughly 85 hours of emotion-neutral recordings spanning across 218 native Chinese mandarin speakers. Their auxiliary attributes such as gender, age group and native accents are explicitly marked and provided in the corpus. Moreover, transcripts in Chinese character-level and pinyin-level are provided along with the recordings. We also present some data processing strategies and techniques which match with the characteristics of the presented corpus and conduct experiments on multiple speech-synthesis systems to assess the quality of the generated speech samples, showing promising results. The corpus is available online at openslr.org/93/ under Apache v2.0 license.

Currently, many multi-speaker speech synthesis and voice conversion systems address speaker variations with an embedding vector. Modeling it directly allows new voices outside of training data to be synthesized. 
GMM based approaches such as Tacospawn are favored in literature for this generation task, but there are still some limitations when difficult conditionings are involved. 
In this paper, we propose VoiceLens, a semi-supervised flow-based approach, to model speaker embedding distributions for multi-conditional speaker generation. VoiceLens maps speaker embeddings into a combination of independent attributes and residual information. It allows new voices associated with certain attributes to be \textit{generated} for existing TTS models, and attributes of known voices to be meaningfully \textit{edited}. We show in this paper, VoiceLens displays an unconditional generation capacity that is similar to Tacospawn while obtaining higher controllability and flexibility when used in a conditional manner. In addition, we show synthesizing less noisy speech from known noisy speakers without re-training the TTS model is possible via solely editing their embeddings with a SNR conditioned VoiceLens model. Demos are available at \url{sos1sos2sixteen.github.io/voicelens}.
\end{abstract}

\section{Introduction}
\label{sec:intro}

Current text-to-speech synthesis and voice conversion systems typically use extra utterance level conditioning vectors to address speaker variances in multi-speaker training, endowing the acoustic model the ability to produce different voices \cite{tan_survey_2021}. The source of this conditioning vector is usually (a) jointly learnt with the acoustic model as a look-up-table (LUT)~\cite{gibiansky_deep_2017}, (b) produced by a speaker encoding module pretrained with speaker verification task \cite{jia_transfer_2018} or jointly optimized with the synthesis model~\cite{wang_style_nodate}\cite{akuzawa_expressive_2019}. If the conditioning vector can be inferred given reference audios of known target speakers, the acoustic model may be used to produce synthesis speech with a similar voice as the reference. This task is studied under zero-shot voice cloning in literature \cite{cooper_zero-shot_2020}\cite{xie_multi-speaker_2021}. 

We, on the other hand, focus on the referenceless case termed \textit{speaker generation} \cite{stanton_speaker_2021-1} in which we aim to generate novel voices that do not correspond to any existing reference audios. Previous attempts at creating hybrid voices in TTS/VC by generating embeddings mainly include linear interpolations~\cite{fang_speaker_2019} or using GMM-based generative models to describe the distribution of the embeddings~\cite{turner_generating_2022}. These methods can be used to create large synthetic corpora with specific textual content and speaker variations as a data-augmentation technique for down-stream speech tasks~\cite{huang_synth2aug_2021}. It can also be leveraged to generate new ``identities'' in a voice-conversion context for speaker annonymizations~\cite{tomashenko_voiceprivacy_2022}. Tacospawn \cite{stanton_speaker_2021-1} builds upon \cite{huang_synth2aug_2021} to use a GMM parameterized by speaker attributes, expanding the task to conditional generation where the sampled speakers' attributes can be specified in advance. However, our experiments show that the simplistic nature of the GMM approach suffers from difficult decision boundaries in the embedding space, resulting in the degradation of attribute controllability under certain scenarios (e.g. child gender specification).

Furthermore, VoiceMe~\cite{van_rijn_voiceme_2022} directly models the speaker embedding with a human-in-the-loop sampling technique~\cite{jacoby_nori_gibbs_2020} to produce specific voices associated with face imagery. Face-voice association in a voice-conversion context has been explored before~\cite{lu_face-based_2021}, but this work employs a \textit{post-hoc} perspective to find the target iteratively by evaluating intermediate synthesis samples after the fact of TTS training.
Concurrent to our work, \cite{bilinski_creating_nodate} proposed a Flow-TTS~\cite{miao_flow-tts_2020} based architecture for creating new voices. Their approach focuses on designing more capable acoustic modeling alternatives to Tacotron while still adopting tacospawn as their embedding generation strategy. 

In this paper, we propose VoiceLens, a semi-supervised flow based generative classifier approach, to model speaker embedding distributions under multiple conditions. 
%We first associate every embedding with $0$ to $N$ discrete or continuous labels of interest in the form of $N$-tuples. 
We first associate every embedding with labels in the form of $N$-tuples, where each value in the label tuple can either be present or absent.
The ratio of annotation depends empirically on the labeling method employed and the difficulty of their associated decision boundaries. Then we bijectively map the TTS embeddings to the flow's base distribution where respective labels are independent.
%, and their values differently distributed. 
%This mapping can be seen as imposing a partially interpretable structure defined by the base distribution on the embeddings.
We show, by sampling from desired conditional distributions of the base variable, novel speaker embeddings associated with specific traits can be generated with strong controllability. Further, the flow-based architecture could perform inference as well as generation. Chaining the two paths of the flow allows us to \textit{edit} traits of existing embeddings. We show, by manipulating the embedding inputs only, it is possible to de-noise some known noisy speakers without modifying the trained syntheiszer under our formulation.
\section{Background}
\subsection{Tacospawn}

Tacospawn \cite{stanton_speaker_2021-1} proposed using $K$-component isotropic GMMs to describe the distribution of the embedding $\mathbf{e}$ under the condition of the speaker metadata $c$ (i.e. gender or accents). The mixture model are parameterized by a dense neural network whose inputs are one-hot encoded metadata $c$. The parameter network is trained along with the main TTS model by maximum likelihood estimation. But gradient propagation between this network and the main TTS model are detached, making the TTS isolated from its training dynamics. 

\begin{equation}
\label{eq-tacospawn}
% %\textstyle
p_{\omega}(\mathbf{e}|\mathbf{c})=\sum_{k}^{K}{\pi_{\omega, k}(\mathbf{c}){\cal N}(\mathbf{e}; \mu_{\omega, k}(\mathbf{c}), \sigma^2_{\omega, k}(\mathbf{c})\mathbf{I})}	
\end{equation}

Because of the categorical nature of the metadata $c$, the parameter network for the conditional GMM is itself essentially a LUT for multiple distinct GMMs estimated during training. This implies that the conditioning works by dividing the training data into multiple one-class subsets. And each GMM learns in a one-class density-estimation fashion. Although this approach performs well for modeling the unconditional distribution $p(\mathbf{e})$~\cite{stanton_speaker_2021-1}~\cite{turner_generating_2022}~\cite{huang_synth2aug_2021}, it results in degraded discrimination between classes with difficult decision boundaries.

\subsection{Normalizing Flow as Generative Classifiers}
Normalizing flow is a type of probabilistic modeling method for describing the complex data distribution $p_\mathbf{e}(\mathbf{e})$. It consist of a base distribution $p_\mathbf{z}(\mathbf{z})$ that is simple to evaluate and perform sampling with, and a bijective mapping $f_\theta: \mathbf{e} \rightarrow \mathbf{z}$ from data to base. 
The mapping $f_\theta$ parameterized by $\theta$ is a neural network designed to be invertible and has a tractable Jacobian $\mathcal{J}_f(\cdot)$~\cite{papamakarios_normalizing_2021}. Such that the exact log-likelihood of $\mathbf{e}$ can be efficiently computed using a change of variable as in eq.\ref{eq_flow}. Thus $\theta$ can be learned from a direct MLE on a set of data samples.

%Normalizing flow is a type of bijective mapping between data from a complex distribution $\mathbf{e}\sim p_\mathbf{e}(\mathbf{e})$ to a base distribution $\mathbf{z}\sim p_\mathbf{z}(\mathbf{z})$ that is simple to evaluate and draw samples from. 

\begin{equation}
\label{eq_flow}
\log p_\mathbf{e}(\mathbf{e}) = \log p_\mathbf{z}(f(\mathbf{e})) + \log|\det\mathcal{J}_f(\mathbf{e})|
\end{equation}

By setting the base distribution to condition on some label $y$, the likelihood in eq.\ref{eq_flow} can simply be re-written for $\log p_\mathbf{e}(\mathbf{e}|y)$. 

\begin{equation}
\label{eq_cls}
\log p_\mathbf{e}(\mathbf{e}|y) = \log p_\mathbf{z}(f(\mathbf{e})|y) + \log|\det\mathcal{J}_f(\mathbf{e})|
\end{equation}

As a result, classification can be done in $\mathbf{z}$ using Bayes theorem (eq.\ref{eq_bayes}). And generation of data from a particular class $y$ can be achieved by first sampling from the corresponding conditional base distribution, then transforming the samples to data space with the reverse path of the flow $f_{\theta}^{-1}$ (eq.\ref{eq_sample}), effectively making the flow a generative classifier.

\begin{figure}[t]
	\centering
	\includegraphics[width=1.05\columnwidth]{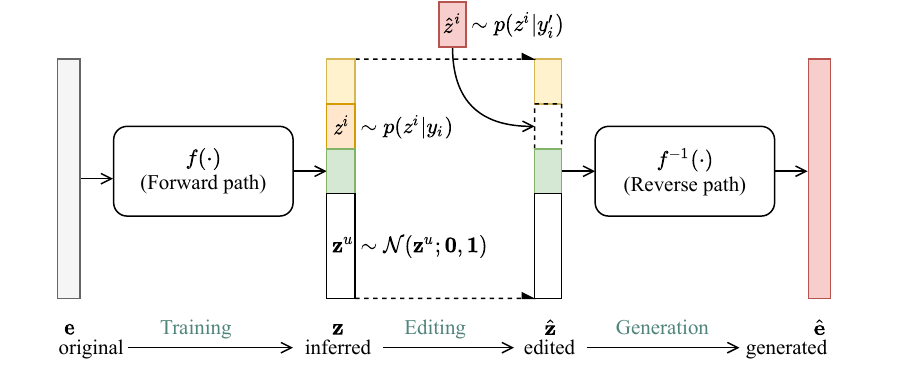}
	\caption{Overview of VoiceLens. The based variable $\mathbf{z}$ is partitioned into multiple sections. Each $z^i$ specialize in representing specific attribute introduced by supervision. The residual $\mathbf{z}^u$ encodes rich variations in speaker characteristics.}
	
%	$\mathbf{z}^u$ encodes rich speaker variations, while $z^i$s specialize in representing specific attributes.}
	\label{fig-spk-edit}
\end{figure}
% \vspace{-1.5em}
\begin{equation}
\label{eq_bayes}
\hat{y}=\mathop{\arg\max}\limits_{y} p_\mathbf{z}(y|f_{\theta}(\mathbf{e})) = \mathop{\arg\max}\limits_{y} p_y(y)p_\mathbf{z}(f_{\theta}(\mathbf{e})|y)
\end{equation}
% \vspace{-1.em}
\begin{equation}
\label{eq_sample}
\mathbf{z}_y\sim p_\mathbf{z}(\mathbf{z}|y); ~~~ \mathbf{e}_y = f_{\theta}^{-1}(\mathbf{z}_y)
\end{equation}

\section{Methods}

VoiceLens models the distribution $p(\mathbf{e}|\mathbf{y})$ of speaker embedding $\mathbf{e}$ conditioned on a set of multi-labels $\mathbf{y}$ by fitting pairs of training data sampled from $p(\mathbf{e}, \mathbf{y})$ to a generative classification flow. The flow is composed of a stack of unconditional bijective transformations and a conditional base distribution $p(\mathbf{z}|\mathbf{y})$. The base distribution is designed such that its likelihood has a closed-form solution for any condition $\mathbf{y}$, allowing maximum likelihood estimation to be performed even for partial or completely marginalized conditions.

Both conditional and unconditional generation can be performed by transforming samples from the appropriate base distribution. Additionally, we could manipulate attributes of existing embeddings by first converting it to $\mathbf{z}$, swapping in new values then converting back to $\mathbf{e}$. We designate this task as \textit{speaker editing} in this paper. An overview of VoiceLens is shown in fig.~\ref{fig-spk-edit}.

\subsection{Supporting Distribution and Post-hoc Labeling}
Assuming independence between labels, the data distribution $p(\mathbf{e},\mathbf{y})$ can be broken down to the unconditional distribution $p(\mathbf{e})$ and multiple classifiers of $y_i$ as $p(\mathbf{e})\prod_i p(y_i|\mathbf{e})$. The collection of train dataset itself can be seen as sampling from this distribution. But training a flow on a few hundred samples that is typical of synthesis corpora leads to unstable performance. As GMM models are generally competent at describing $p(\mathbf{e})$, we approximate it with a supporting GMM $p_\phi(\mathbf{e})$ estimated from known speakers to provide additional regularization to the underlying structure of $\mathbf{e}$ for our flow. 

New embeddings without label information can be acquired by sampling $p_\phi(\mathbf{e})$. In such cases, we either leave $y_i$ empty for semi-supervised learning, or approximate the classifiers of $y_i$ on $\mathbf{e}$ post-hoc with existing classifiers of speech signals. The speech signals being classified are synthesized by conditioning TTS on the newly sampled $\mathbf{e}$. This post-hoc labeling procedure has the additional benefit of including the exact TTS system being modeled into consideration. This inclusion is desirable since TTS cannot achieve perfect reconstruction, and certain high-level traits such as speaker noise level and reading pace may deviate from the ground truths, thus can only be captured post-hoc.

\subsection{Multi-label Flow for Speaker Generation}
\label{sec-base-dist}
Data from $p(\mathbf{e},\mathbf{y})$ consist of pairs of speaker embedding $\mathbf{e}$ and the corresponding multi-label $\mathbf{y}$.For each $\mathbf{e}_i \in \mathbb{R}^d$,$\mathbf{y}_i$ contains multiple attributes, e.g. in our experiments, a (\mbox{gender}, \mbox{age-group}, \mbox{SNR}) triplet. Each attribute  takes a value from its corresponding label-set $L^i$. $L^i$ can be either continuous or categorical, and augmented with a special symbol indicating empty value. That is, the $l$-tuple label $\mathbf{y}$ of each embedding $\mathbf{e}$ belongs to the Cartesian product $\prod_i^l L^i\cup\{\varnothing\}$, where $\varnothing$ denotes unknown values. For example, $(\textbf{F}, \varnothing, \varnothing)$ may label any embedding for a female-like voice regardless of its age-group and SNR level. 

A normalizing flow denoted by $f_\theta$ bijectively maps $\mathbf{e}$ to $\mathbf{z}$ of the same dimensionality, transforming the complex distribution $p(\mathbf{e}|\mathbf{y})$ to a simpler base distribution $p(\mathbf{z}|\mathbf{y})$. 
%The parameters of the flow model are learned using direct MLE on the training data by a change of variable from $\mathbf{e}$ to the base variable $\mathbf{z}$.
The base distribution is conventionally chosen to be diagonal Gaussian, resulting in the disentanglement (independently distributed) of every dimensions in $\mathbf{z}$.
As a slightly relaxed assumption, we may partition~\cite{fetaya_understanding_2020} the base variable into $l+1$ independent sections as $\mathbf{z}:=[\mathbf{z}^{1},..., \mathbf{z}^{l}, \mathbf{z}^{u}]$, each corresponding to one of the $l$ attributes, plus one additional unconditional section. This gives the factorization of conditions in eq.\ref{eq_fac}.

\begin{equation}
\label{eq_fac}
%\textstyle
p(\mathbf{z}|\mathbf{y})=p(\mathbf{z}^u)\prod_i^l p(\mathbf{z}^{i}|y_i)
\end{equation}

We choose each term in eq.\ref{eq_fac} to be isotropic Gaussian with mean conditioned on their respective labels. 

\begin{equation}
\label{eq_cond}
%%\textstyle
p(\mathbf{z}^i|y_i) = \mathcal{N}(\mathbf{z}^i; \mathbf{\mu}(y_i), \mathbf{I}) ~~~~~~\\
p(\mathbf{z}^u) = \mathcal{N}(\mathbf{z}^u; \mathbf{0}, \mathbf{I})
\end{equation}

%\begin{equation}
%\label{eq_uncond}
%
%\end{equation}

For categorical attribute $L^i$, the mean vector $\mathbf{\mu}(y_i)$ is a look-up-table for each element of $L^i$. As an example, we choose the center ($\mathbf{0}$) and a shifted value \cite{csiszarik_negative_2022} ($\alpha\mathbf{I}$) to be the mean vectors of the $2$-class gender attribute $\{\textbf{F}, \textbf{M}\}$. For continuous attribute such as SNR estimations in decibel. We choose the mean to be a linear function of $y_i$.

The log-likelihood of a data pair $(\mathbf{z}, \mathbf{y})$ can be calculated using eq.\ref{eq_fac} for $\mathbf{y}$s with full conditions (no $\varnothing$s). For partial $\mathbf{y}$s, assuming the $i$-th attribute is omitted (set to $\varnothing$), the likelihood is calculated from the marginal distribution $p(\mathbf{z}|\mathbf{y}_{-i})$. This marginal is evaluated the same as in eq.\ref{eq_fac} except for the $i$-th conditional, now in the form of eq.\ref{eq_margin_cat} or eq.\ref{eq_margin_cont} depending on the type of the label-set $L^i$. 

\begin{equation}
\label{eq_margin_cat}
%\textstyle
p(\mathbf{z}^{i}|\varnothing) := p(\mathbf{z}^{i}) = \sum_{y\in L^i}p(y)p(\mathbf{z}^i|y)
\end{equation} 

\begin{equation}
\label{eq_margin_cont}
%\textstyle
p(\mathbf{z}^i|\varnothing) := p(\mathbf{z}^i) = \int_{y\in L^i}p(y)p(\mathbf{z}^i|y)dy
\end{equation}

The marginalized form in eq.\ref{eq_margin_cat} effectively becomes an isotropic Gaussian mixture for categorical $y_i$. As for the marginalized distribution for continuous $y_i$ in eq.\ref{eq_margin_cont}. By choosing $y\sim \mbox{uniform}(a,b)$, it becomes a Bhattacharje distribution\cite{bhattacharjee_dimensional_1963}, whose pdf can be evaluated via the the standard Gaussian cdf. 
It's clear that the above procedure can be generalized to $\mathbf{y}$s with any number of $\varnothing$s. So the likelihood function described by eq.[\ref{eq_fac},\ref{eq_margin_cat},\ref{eq_margin_cont}] can be used as the base likelihood for a semi-supervised multi-conditional normalizing flow, where training data may only be partially labeled.

\subsection{Semi-supervised Training}
Using the base distribution described in section \ref{sec-base-dist}, a flow-based generative classifier can be trained in a semi-supervised manner~\cite{izmailov_semi-supervised_2020}~\cite{atanov_semi-conditional_2020}. 
The MLE loss w.r.t $\theta$ for a mini-batch of $N$ data pairs $\{(\mathbf{e}^j, \mathbf{y}^j)\}^N$ where unlabeled $y_i$s are marginalized is given by eq.\ref{eq_loss_mle}.

\begin{equation}
\label{eq_loss_mle}
%\textstyle
\mathcal{L}_{\tiny{\mbox{MLE}}}=-\sum_j^N[\log p_\mathbf{z}(f_\theta(\mathbf{e}^j)|\mathbf{y}^j) + \log|\det\mathcal{J}_f(\mathbf{e}^j)|]
\end{equation}

The loss function in eq.\ref{eq_loss_mle} is further regularized using the consistency regularization method proposed in \cite{izmailov_semi-supervised_2020}. 
Assuming the true classification is robust against small perturbations in data, we assert this belief by adding regularization data $\{\hat{\mathbf{e}}^j\}^{M}$ drawn from an unconditional generative model for $\mathbf{e}$. The associate labels $\{ \hat{\mathbf{y}}^j \}$ are predicted from the perturbed data $\{\hat{\mathbf{e}}^j + \epsilon\}^{M}$ according to eq.\ref{eq_bayes}. Let the log-likelihood of the regularization data be denoted $\mathcal{L}_{\tiny{\mbox{REG}}}$. The final loss term of the semi-supervised flow is given by $\mathcal{L} = \mathcal{L}_{\tiny{\mbox{MLE}}} + \mathcal{L}_{\tiny{\mbox{REG}}} $.

%\begin{equation}
%\label{eq_loss_all}
%\mathcal{L} = \mathcal{L}_{\tiny{\mbox{MLE}}} + \mathcal{L}_{\tiny{\mbox{REG}}} 
%\end{equation}

%\subsection{Evalutaion Methods}
%We evaluate the performance of the proposed approach by its generation capability and controllability. As we employ a supporting GMM to regularize the structure of the overall distribution, we expect on par unconditional performance with GMM models as measured by naturalness MOS scoring and the speaker generation metrics in \cite{stanton_speaker_2021-1}. We also report the approximated clique-number $\omega(G)$ of the speaker exclusivity graph proposed in~\cite{shi_aishell-3_2021} to provide intuitive measure on the richness of the potential voices. 
%The controllability of discrete labels in the generation process is measured by its control-label accuracy or linear regression with respect to human and automatic post-hoc judgements. 
\section{Experimental Results}
\subsection{Experimental Setup}
\subsubsection{Supporting Environments}
% first data, then model
We performed experiments on DiDiSpeech-2~\cite{guo_didispeech_2021}.
%a public speech corpus containing around \textcolor{black}{227} hours of \textcolor{black}{48kHz} speech from \textcolor{black}{1500} speakers. The speakers span the age from \textcolor{black}{6} to \textcolor{black}{60}. 
The recording was done in various conditions, resulting in wide SNR levels distributed within \textcolor{black}{[20, 60]} dB. 
For age-groups, we employed a simple \textit{child-adult} two-part division where speakers under \textcolor{black}{13} are considered as \textit{child}.
A subset of \textcolor{black}{1489} speakers with more than \textcolor{black}{100} utterances were selected for TTS training.
%Train-test partition was done by reserving \textcolor{black}{10} random utterances from each speaker for testing.
%\footnote{\url{github.com/jaywalnut310/vits}} 
We used the official VITS \cite{kim_conditional_2021} implementation with a $256$ dimensional speaker LUT as the base multi-speaker TTS system. 
%The model was trained from scratch to $1250$k steps with a batch-size of \textcolor{black}{$256$}.
We used the pre-trained speaker recognition model provided by the \url{speechbrain} toolkit \cite{ravanelli_speechbrain_2021} as the speaker feature encoder in our evaluations. A similar architecture with an additional gender classification branch was trained with DiDiSpeech as the gender classifier for post-hoc labeling in our experiments.

\subsubsection{Generation Models}
% first data, then model
Trained speaker embeddings from the VITS model were extracted as the basis for speaker generation modeling. We performed train-test partition on the embedding data by reserving \textcolor{black}{150} randomly selected speakers for validation.
%Note this is different from the train-test partition for TTS, which was done in terms of utterances. 
All speaker generative models in our experiments were trained only on this subset of TTS embeddings.
%All speakers are present in the TTS test-set for generation evaluation. 

The baseline Tacospawn network was implemented as a dense ReLU network with one-hot encoded (gender, age-group) labels as inputs, which in turn parameterizes a 10 component GMM as in eq.\ref{eq-tacospawn}.
%The output of this network was used to parameterize a \textcolor{black}{10} component isotropic Gaussian mixture as in eq.\ref{eq-tacospawn}. 

We implemented~\cite{durkan_nflows_2020} the proposed flow transform as a sequence of 5 masked affine autoregressive flows~\cite{papamakarios_masked_2018}. The base distribution partitions the $256$ dimensional $\mathbf{z}$ into four parts of size $(1, 1, 1, 253)$ for modeling the gender, age-group, SNR and residual information respectively. For $2$-class discrete labels (gender and age-group) we set the shift $\alpha=6$. And for the continuous case, we used raw SNR value within range $[25, 55]$ dB as the mean of the base distribution~($\mu = y_{\tiny\mbox{snr}}$). We used embeddings of known speakers as well as samples from supporting GMMs to train our proposed flow till validation likelihood stops improving.
In preliminary experiments, we've identified child gender distinction to be a difficult case for GMMs, whereas the age-group distinction can be easily modeled. Therefore, for supporting distributions, we modeled data from different age-groups with separate GMMs and only used strong post-hoc gender labels in the child subset. We also used a separate unconditional GMM to produce samples with only post-hoc SNR labels. Note, throughout our experiments on VoiceLens, labels \textit{always} contain partial values. The semi-supervised setup allows us to use partially labeled data to model the full conditions described by $p(\mathbf{z}|\mathbf{y})$.

\subsection{Speaker Generation}

\begin{table}
\centering
\caption{Unconditional Speaker Generation Results}
\begin{tabular}{crrrrc} 
\toprule
\multirow{2}{*}{System}       & \multicolumn{4}{c}{Speaker distance~\cite{stanton_speaker_2021-1}}                                                                     & Naturalness  \\
                              & \multicolumn{1}{l}{$s2s$} & \multicolumn{1}{l}{$s2g$} & \multicolumn{1}{l}{$g2g$} & \multicolumn{1}{l}{$s2t$-$s$} & MOS ($\pm95\%$ CI)    \\ 
%\cmidrule{2-6}
\midrule
%\multicolumn{1}{l}{synthesized} & \multirow{3}{*}{N/A}    & \textcolor{red}{$3.6\pm.1$}        \\
\multicolumn{1}{l}{recording} & \multicolumn{4}{c}{\multirow{2}{*}{/}}    &                        	\textcolor{black}{$3.44\pm0.10$} \\
\multicolumn{1}{l}{synthesized} & \multicolumn{4}{l}{}  &                        					\textcolor{black}{$3.56\pm0.09$}\\
\multicolumn{1}{l}{tacospawn} & $0.26$   & $0.26$       & $0.22$    & \textcolor{black}{$0.07$}    & \textcolor{black}{$3.59\pm0.09$}        \\
\multicolumn{1}{l}{voicelens~} & $0.26$   & $0.26$       & $0.22$    & $0.07$    					& \textcolor{black}{$3.52\pm0.10$}        \\
\bottomrule
\end{tabular}
%\begin{tablenotes}
%	\centering
%    \item[1]{$^*$ MOS differences in the last 3 entries are insignificant.}
%\end{tablenotes}
\label{tab-uncond}
\end{table}

\begin{figure}[t]
	\centering
	\includegraphics[width=1.0\columnwidth]{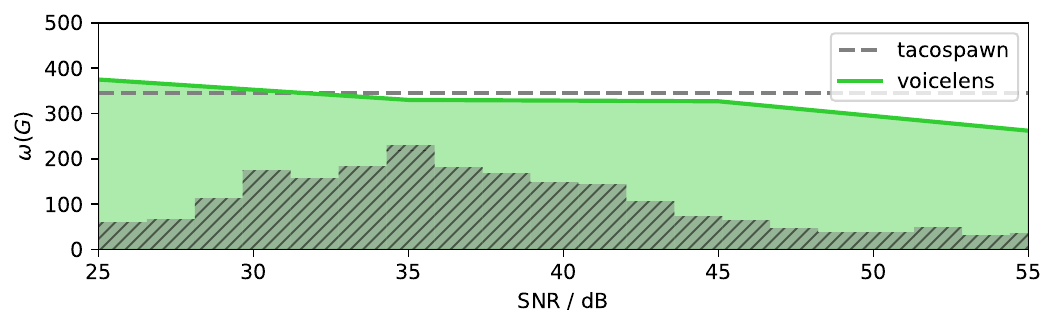}
	% \vspace{-1.5em}
	\caption{plot of $\omega(G)$ estimations by SNR level.
	Hatched bars depicts a histogram of per speaker SNR level of train speakers from DiDiSpeech.
%	\textbf{Bottom scale:} colored region depicts $\omega(G)$ of proposed by SNR level, dashed line shows $\omega(G)$ for tacospawn. \textbf{Top scale: } hatched bars show a histogram of \#train-speakers by SNR.
}
	\label{fig-clique}
\end{figure}

%The speaker distance metrics~\cite{stanton_speaker_2021-1} measured from 1:1 generated populations and naturalness MOS scores are reported in table~\ref{tab-uncond}. 

The speaker distance metrics ($s2s$, $g2s$, etc.)~\cite{stanton_speaker_2021-1} reflect how close the statistics of generated speakers match that of the real speakers
% in terms of speaker recognition model we employed, 
 and were used as an objective measure of generation performance in~\cite{stanton_speaker_2021-1}. In general, $g2s$ and $g2g$ equaling $s2s$ indicates effective speaker generation. We also use subjective MOS to test the validity of the generated voices. 
The MOS were obtained from a random subset of \textcolor{black}{15} utterances for each system, and scored by \textcolor{black}{19} native mandarin speakers. 
The distance and MOS results are reported in table-\ref{tab-uncond}.
We also report the approximated clique-number $\omega(G)$~\cite{shi_aishell-3_2021} of Tacospawn and VoiceLens. 
The $\omega(G)$ reflects the number of speakers that are mutually at least $d$ cos-distance away, hence serves as a estimate of the number of distinct speakers within a pool of generated voices, which is indicative of the \textit{abundance} of novel voices under each condition.
The threshold $d$ was chosen as the $s2s$ value (the typical distance between nearby speakers~\cite{stanton_speaker_2021-1}).
We estimate the $\omega(G)$ value for both systems.
SNR controlling unique to VoiceLens was handled by grouping 10 dB uniform intervals as conditions.
The results are reported in fig.\ref{fig-clique}
%we estimate the $\omega(G)$ value of 500 generated speakers from VoiceLens. We also include a unconditional estimation result from Tacospawn as it cannot be conditioned on 

% We sampled 500 embeddings for $\omega(G)$ estimation. SNR controlling unique to  VoiceLens was handled by grouping 10 dB uniform intervals as conditions. 
%  of the respective gender. 
% We differentiated by genders because female speakers are generally closer together than males in verification systems, this phenomenon was reported in \cite{turner_generating_2022}.

Both objective and subjective evaluation results in table~\ref{tab-uncond} suggest VoiceLens has similar unconditional capabilities to Tacospawn, as the MOS difference between the two systems are insignificant ($p=.29$). This claim is further supported by fig.\ref{fig-clique} as 
the $\omega(G)$ estimates of VoiceLens (colored region) roughly conincides with that of Tacospawn (dashed line).
%colored region (VoiceLens) roughly coincides with the dashed line (Tacospawn).
This result is expected since our flow model was trained with supporting GMMs as $p(\mathbf{e}, \mathbf{y})$. Therefore, they share the same structure in the marginal case. In addition, by sampling unconditionally $5000$ embeddings from VoiceLens, we estimated an $\omega(G)$ of $2858$ ($>1489$). 
%The greater number suggests the existence of novel speakers. 
The larger number provides clear evidence of the existence of novel speakers.

\subsection{Controllability}

%\begin{table}[t]
%\centering
%\caption{Generation Controllability Results}
%\begin{tabular}{lrrll} 
%\toprule
%\multirow{2}{*}{System} & \multicolumn{2}{c}{\textit{child} gender ACC/\%}                    & \multicolumn{2}{c}{SNR control}           \\
%                        & \multicolumn{1}{l}{Classifier} & \multicolumn{1}{l}{Human} & $r$                     & p-value         \\ 
%\midrule
%real-speakers           & \textcolor{Green}{$^*97.78$}   & \textcolor{red}{$90$}                        & \multicolumn{2}{c}{\multirow{3}{*}{/}}  \\
%tacospawn               & \textcolor{Green}{$82.76$}     & \textcolor{red}{$69$}                        & \multicolumn{2}{c}{}                      \\
%separate flows          & \textcolor{Green}{$77.10$}     & \textcolor{red}{$70$}                        & \multicolumn{2}{c}{}                      \\
%\textbf{proposed}       & \textcolor{Green}{$\mathbf{92.70}$}    & \textcolor{red}{$\mathbf{86}$}       & \multicolumn{1}{r}{$\mathbf{0.943}$} & $\mathbf{<10^{-5}}$        \\
%\bottomrule
%\end{tabular}
%\begin{tablenotes}
%	\centering
%    \item[1]{$^*$ only \textcolor{Green}{315} real speakers in \textit{child} subset}
%\end{tablenotes}
%\label{tab-control}
%\end{table}

\begin{table}[t]
\centering
\caption{Generation Controllability Results}
\begin{tabular}{lrrr} 
\toprule
\multirow{2}{*}{System} & \multicolumn{1}{c}{\textit{child} gender/\%}                    & \multicolumn{2}{c}{SNR control}           \\
                        & \multicolumn{1}{c}{ACC/\%}  & $r$                     & p-value         \\ 
\midrule
real-speakers           & \textcolor{black}{$^*97.78$}                        & \multicolumn{2}{c}{\multirow{3}{*}{/}}  \\
tacospawn               & \textcolor{black}{$82.76$}                           & \multicolumn{2}{c}{}                      \\
separate flows          & \textcolor{black}{$77.10$}                     & \multicolumn{2}{c}{}                      \\
\textbf{voicelens}       & \textcolor{black}{$\mathbf{92.70}$}        & \multicolumn{1}{r}{$\mathbf{0.943}$} & $\mathbf{<10^{-5}}$        \\
\bottomrule
\end{tabular}
\begin{tablenotes}
	\centering
    \item[1]{$^*$ only \textcolor{black}{315} real speakers in \textit{child} subset}
\end{tablenotes}
\label{tab-control}
\end{table}

%\begin{figure*}[t]
%	\centering
%	\includegraphics[width=\textwidth]{diagrams-Page-4.drawio.pdf}
%	\caption{Overview of the proposed method. $\mathbf{z}^u$ encodes rich speaker variations, while $z^i$s specialize in representing specific attributes.}
%	\label{fig-exp}
%\end{figure*}

For gender controllability, we generated for each system, $5000$ embeddings with uniform gender conditions from the \textit{child} subset as trials to calculate their respective gender classification accuracy. 
%Each embedding is synthesized with 3 sentences and the classification scores were averaged to obtain the final score for each embedding. 
%A subset of \textcolor{red}{$50$} generated speakers for each system were sampled to perform subjective gender classification. For each speaker, human participants are expected to respond either \textit{male} or \textit{female}. 
The results are reported in table \ref{tab-control}. The additional baseline \textit{separate flow} serves as an ablation study of our approach. It uses separate unconditional flows with identical architecture and training data to the proposed method, but was trained separately for each gender, mimicking the modeling strategy of the Tacospawn baseline. The accuracy results show that VoiceLens attained stronger agreement between the generation \textit{intention} and the perceived \textit{result} obtained from synthesized voices. The lack of controllability in \textit{separate flow} also suggests the proposed generative classification approach's effectiveness in modeling these conditions.

For SNR controllability, we unconditionally generated \textcolor{black}{$500$} speakers and estimated their true SNR levels post-hoc. The Pearson correlation result $r$ in table \ref{tab-control} and fig.\ref{fig-edit}(a) suggest strong correlation between the control variable and the generated speakers' actual SNR level. 
The relation between SNR control and generation capabilities are further explored in fig.\ref{fig-clique}. Though voices with the abundance that roughly matches the unconditional baseline were generated at every SNR level, we observed a slow decreasing trend in $\omega(G)$ when setting higher SNR levels in generation. This phenomenon suggests cleaner voices are generally less plentiful in our experimental generation system. We conjecture this exposes a limitation of the synthesis model itself, as this downward slop can also be observed on the per-speaker SNR histogram of train-set speakers (hatched region in fig.\ref{fig-clique}). 

%\vspace{-0.5em}
\subsection{Known Speaker \textcolor{black}{De-nosing} via Speaker Editing}

\begin{figure}[t]
	\centering
	\includegraphics[width=1.0\columnwidth]{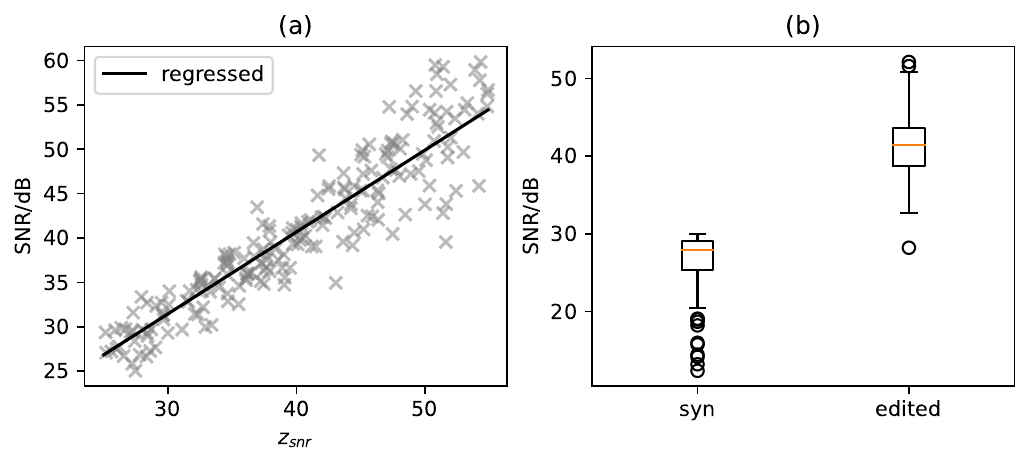}
	\caption{SNR controllability. (a) Scatter plot and linear regression for the control variable $z^{snr}$ and generated voices' SNR. (b) Boxplot showing post-hoc SNR  boosting by speaker editing. }
	\label{fig-edit}
\end{figure}

Speaker editing can be used to manipulate any labeled attributes. However, as gender and age flipping lacks quantifiable metrics. In this section, we focus on post-hoc SNR level manipulation of existing speakers as a case study.
%Multi-speaker TTS systems tend to preserve undesired attributes such as noisy recording conditions from the training data. 
We selected $199$ known speakers from the VITS's speaker LUT, whose synthesis results have lower than 30dB SNR. 
Their embeddings were then edited by increasing their inferenced $z^{snr}$ by $15$ following the editing procedure in fig.\ref{fig-spk-edit}.

A boxplot of the SNR levels of the \textit{original} and \textit{edited} voices are shown in fig.\ref{fig-edit}. The edited speakers gain an average SNR boost of 14.5dB. 
Additionally, the original and edited speakers have an cos-distance median of $0.09$, within the distance tolerance given by $s2s$ in table~\ref{tab-uncond}. We also conducted speaker similarity MOS tests before and after the manipulation. 
%The scores were obtained from $100$ random pairs of either (recording, synthesized) or (recording, edited) stimuli, and assessed by \textcolor{black}{20} native speakers. 
%The MOS study shows a relative similarity loss of \textcolor{black}{$0.54\pm0.09$} after editing, with 30 voices shows no significant loss of similarity. This reading shows that the edited voices profess a general like
We observed on average a relative similarity loss of $0.54\pm0.09$. The per-speaker	 differences in similarity are statistically insignificant in $60\%$ of the cases, showing, albeit non-uniform, the disentanglement of the voice and the contingent features~(noise) in the un-modified multi-speaker TTS model. The proposed VoiceLens is able to capture this pre-existing property and expose a meaningful interface for editing.
%the consistency of voices through the editing process. 

%\textcolor{Green}{Fig-\ref{fig-edit}} shows the mel-spectrogram of a noisy train recording from a known speaker at \textcolor{red}{19dB}. 
%the synthesized utterances at \textcolor{red}{20dB} using embeddings from the LUT table are displayed in \textcolor{red}{fig 2(b)}. 

\section{Conclusions}

This paper proposes VoiceLens, a flow-based generative classifier approach to speaker generation. VoiceLens achieves similar generative capacity to Tacospawn. It is also capable of modeling multiple parallel control labels with partially labeled dataset through a semi-supervised learning process, and extends to applications such as editing. We show in our experiments the greater controllability and flexibility of the proposed approach. 

Multi-speaker TTS systems tend to entangle speaker identity and other contingent features into its learned embeddings. 
One may read VoiceLens as forcing an \textit{a priori} structure onto this embedding space, making navigation within it in a meaningful way possible. 
As a case study, we demonstrate synthesizing de-noised versions of selected noisy speakers post-hoc by solely manipulating the speaker embeddings. Moreover, the choice of the prior also influences our capability of navigation. In our preliminary studies, modeling \textit{gender} as a continuous gradient allowed us to describe gender-neutral voices instead of hard 2-way decisions. 
Nevertheless, the residual information distilled by the proposed method contains characteristic informations that are difficult to be conceptualized into labels. Learning to navigate this residual remains a challenge to full-fledged speaker editing.

%2. limitation analysis, methodological concerns, implications to zero-shot speaker modeling.
%This $2$-class model for gender classification is a simplification of rich diversities involved in how we perceive gender characteristics from speech. We use the $2$-class model in our experiments as a case study to show the proposed methods advantage in modeling difficult decision boundaries. In preliminary experiments, we also modeled gender as a 1-d continuous gradient guided by the gender-classifier's normalized score. This alternate model gave us the ability to describe gender-neutral voices in our system. 

%3. general conclusion.

%\clearpage 
%\vfill \pagebreak

\bibliographystyle{IEEEbib}
\bibliography{simple}

\begin{thebibliography}{10}

\bibitem{tan_survey_2021}
Xu~Tan, Tao Qin, Frank Soong, and Tie-Yan Liu,
\newblock ``A {Survey} on {Neural} {Speech} {Synthesis},''
\newblock {\em arXiv preprint arXiv:2106.15561}, July 2021.

\bibitem{gibiansky_deep_2017}
Andrew Gibiansky, Sercan Arik, Gregory Diamos, John Miller, Kainan Peng, Wei
  Ping, Jonathan Raiman, and Yanqi Zhou,
\newblock ``Deep {Voice} 2: {Multi}-{Speaker} {Neural} {Text}-to-{Speech},''
\newblock in {\em Advances in {NIPS}}, 2017, vol.~30.

\bibitem{jia_transfer_2018}
Ye~Jia, Yu~Zhang, Ron Weiss, Quan Wang, Jonathan Shen, Fei Ren, zhifeng Chen,
  Patrick Nguyen, Ruoming Pang, Ignacio Lopez~Moreno, and Yonghui Wu,
\newblock ``Transfer {Learning} from {Speaker} {Verification} to {Multispeaker}
  {Text}-{To}-{Speech} {Synthesis},''
\newblock in {\em Advances in {NIPS}}, 2018, vol.~31.

\bibitem{wang_style_nodate}
Yuxuan Wang, Daisy Stanton, Yu~Zhang, R.~J. Skerry{-}Ryan, Eric Battenberg,
  Joel Shor, Ying Xiao, Ye~Jia, Fei Ren, and Rif~A. Saurous,
\newblock ``Style tokens: Unsupervised style modeling, control and transfer in
  end-to-end speech synthesis,''
\newblock in {\em Proceedings of {ICML}}, 2018, vol.~80, pp. 5167--5176.

\bibitem{akuzawa_expressive_2019}
Kei Akuzawa, Yusuke Iwasawa, and Yutaka Matsuo,
\newblock ``Expressive speech synthesis via modeling expressions with
  variational autoencoder,''
\newblock in {\em Proceedings of Interspeech}, 2018, pp. 3067--3071.

\bibitem{cooper_zero-shot_2020}
Erica Cooper, Cheng{-}I Lai, Yusuke Yasuda, Fuming Fang, Xin Wang, Nanxin Chen,
  and Junichi Yamagishi,
\newblock ``Zero-shot multi-speaker text-to-speech with state-of-the-art neural
  speaker embeddings,''
\newblock in {\em Proceedings of {ICASSP}}, 2020, pp. 6184--6188.

\bibitem{xie_multi-speaker_2021}
Qicong Xie, Xiaohai Tian, Guanghou Liu, Kun Song, Lei Xie, Zhiyong Wu, Hai Li,
  Song Shi, Haizhou Li, Fen Hong, Hui Bu, and Xin Xu,
\newblock ``The {Multi}-{Speaker} {Multi}-{Style} {Voice} {Cloning} {Challenge}
  2021,''
\newblock in {\em Proceedings of {ICASSP}}, June 2021, pp. 8613--8617.

\bibitem{stanton_speaker_2021-1}
Daisy Stanton, Matt Shannon, Soroosh Mariooryad, R.~J. Skerry{-}Ryan, Eric
  Battenberg, Tom Bagby, and David Kao,
\newblock ``Speaker generation,''
\newblock in {\em Proceedings of {ICASSP}}, 2022, pp. 7897--7901.

\bibitem{fang_speaker_2019}
Fuming Fang, Xin Wang, Junichi Yamagishi, Isao Echizen, Massimiliano Todisco,
  Nicholas Evans, and Jean-Francois Bonastre,
\newblock ``Speaker {Anonymization} {Using} {X}-vector and {Neural} {Waveform}
  {Models},''
\newblock {\em arXiv preprint arXiv:1905.13561}, May 2019.

\bibitem{turner_generating_2022}
Henry Turner, Giulio Lovisotto, and Ivan Martinovic,
\newblock ``Generating identities with mixture models for speaker
  anonymization,''
\newblock {\em Computer Speech \& Language}, vol. 72, pp. 101318, 2022.

\bibitem{huang_synth2aug_2021}
Yiling Huang, Yutian Chen, Jason Pelecanos, and Quan Wang,
\newblock ``{Synth2Aug}: {Cross}-{Domain} {Speaker} {Recognition} with {TTS}
  {Synthesized} {Speech},''
\newblock in {\em Proceedings of {IEEE} {SLT}}, 2021, pp. 316--322.

\bibitem{tomashenko_voiceprivacy_2022}
Natalia~A. Tomashenko, Xin Wang, Emmanuel Vincent, Jose Patino, Brij Mohan~Lal
  Srivastava, Paul{-}Gauthier No{\'{e}}, Andreas Nautsch, Nicholas W.~D. Evans,
  Junichi Yamagishi, Benjamin O'Brien, Ana{\"{\i}}s Chanclu,
  Jean{-}Fran{\c{c}}ois Bonastre, Massimiliano Todisco, and Mohamed Maouche,
\newblock ``The voiceprivacy 2020 challenge: Results and findings,''
\newblock {\em Comput. Speech Lang.}, vol. 74, pp. 101362, 2022.

\bibitem{van_rijn_voiceme_2022}
Pol van Rijn, Silvan Mertes, Dominik Schiller, Piotr Dura, Hubert Siuzdak,
  Peter M.~C. Harrison, Elisabeth Andr{\'{e}}, and Nori Jacoby,
\newblock ``Voiceme: Personalized voice generation in {TTS},''
\newblock in {\em Proceedings of Interspeech}, 2022, pp. 2588--2592.

\bibitem{jacoby_nori_gibbs_2020}
Peter M.~C. Harrison, Raja Marjieh, Federico Adolfi, Pol van Rijn, Manuel
  Anglada{-}Tort, Ofer Tchernichovski, Pauline Larrouy{-}Maestri, and Nori
  Jacoby,
\newblock ``Gibbs sampling with people,''
\newblock in {\em Advances in NeurIPS}, 2020.

\bibitem{lu_face-based_2021}
Hsiao{-}Han Lu, Shao{-}En Weng, Ya{-}Fan Yen, Hong{-}Han Shuai, and Wen{-}Huang
  Cheng,
\newblock ``Face-based voice conversion: Learning the voice behind a face,''
\newblock in {\em {ACM} Multimedia Conference}, 2021, pp. 496--505.

\bibitem{bilinski_creating_nodate}
Piotr Bilinski, Tom Merritt, Abdelhamid Ezzerg, Kamil Pokora, Sebastian Cygert,
  Kayoko Yanagisawa, Roberto Barra-Chicote, and Daniel Korzekwa,
\newblock ``Creating new voices using normalizing flows,''
\newblock in {\em Proceedings of Interspeech}, 2022.

\bibitem{miao_flow-tts_2020}
Chenfeng Miao, Shuang Liang, Minchuan Chen, Jun Ma, Shaojun Wang, and Jing
  Xiao,
\newblock ``Flow-{TTS}: {A} {Non}-{Autoregressive} {Network} for {Text} to
  {Speech} {Based} on {Flow},''
\newblock in {\em Proceedings of {ICASSP}}, 2020, pp. 7209--7213.

\bibitem{papamakarios_normalizing_2021}
George Papamakarios, Eric~T. Nalisnick, Danilo~Jimenez Rezende, Shakir Mohamed,
  and Balaji Lakshminarayanan,
\newblock ``Normalizing flows for probabilistic modeling and inference,''
\newblock {\em J. Mach. Learn. Res.}, vol. 22, pp. 57:1--57:64, 2021.

\bibitem{fetaya_understanding_2020}
Ethan Fetaya, J{\"{o}}rn{-}Henrik Jacobsen, Will Grathwohl, and Richard~S.
  Zemel,
\newblock ``Understanding the limitations of conditional generative models,''
\newblock in {\em Proceedings of {ICLR}}, 2020.

\bibitem{csiszarik_negative_2022}
Adri{\'a}n Csisz{\'a}rik, Beatrix Benk{\H{o}}, and D{\'a}niel Varga,
\newblock ``Negative sampling in variational autoencoders,''
\newblock in {\em Proceedings of IEEE CITDS}, 2022, pp. 63--68.

\bibitem{bhattacharjee_dimensional_1963}
GP~Bhattacharjee, SNN Pandit, and R~Mohan,
\newblock ``Dimensional chains involving rectangular and normal
  error-distributions,''
\newblock {\em Technometrics}, vol. 5, no. 3, pp. 404--406, 1963.

\bibitem{izmailov_semi-supervised_2020}
Pavel Izmailov, Polina Kirichenko, Marc Finzi, and Andrew~Gordon Wilson,
\newblock ``Semi-{Supervised} {Learning} with {Normalizing} {Flows},''
\newblock in {\em Proceedings of ICML}, 2020, pp. 4615--4630.

\bibitem{atanov_semi-conditional_2020}
Andrei Atanov, Alexandra Volokhova, Arsenii Ashukha, Ivan Sosnovik, and Dmitry
  Vetrov,
\newblock ``Semi-{Conditional} {Normalizing} {Flows} for {Semi}-{Supervised}
  {Learning},''
\newblock {\em arXiv preprint arXiv:1905.00505}, June 2020.

\bibitem{guo_didispeech_2021}
Tingwei Guo, Cheng Wen, Dongwei Jiang, Ne~Luo, Ruixiong Zhang, Shuaijiang Zhao,
  Wubo Li, Cheng Gong, Wei Zou, Kun Han, and Xiangang Li,
\newblock ``Didispeech: {A} large scale mandarin speech corpus,''
\newblock in {\em Proceedings of ICASSP}, 2021, pp. 6968--6972.

\bibitem{kim_conditional_2021}
Jaehyeon Kim, Jungil Kong, and Juhee Son,
\newblock ``Conditional {Variational} {Autoencoder} with {Adversarial}
  {Learning} for {End}-to-{End} {Text}-to-{Speech},''
\newblock in {\em Proceedings of ICML}, 2021, pp. 5530--5540.

\bibitem{ravanelli_speechbrain_2021}
Mirco Ravanelli, Titouan Parcollet, Peter Plantinga, Aku Rouhe, Samuele
  Cornell, Loren Lugosch, Cem Subakan, Nauman Dawalatabad, Abdelwahab Heba, and
  Jianyuan Zhong{, et al.},
\newblock ``{SpeechBrain}: A general-purpose speech toolkit,''
\newblock {\em arXiv preprint arXiv:2106.04624}, 2022.

\bibitem{durkan_nflows_2020}
Conor Durkan, Artur Bekasov, Iain Murray, and George Papamakarios,
\newblock ``nflows: normalizing flows in {PyTorch},'' 2020.

\bibitem{papamakarios_masked_2018}
George Papamakarios, Iain Murray, and Theo Pavlakou,
\newblock ``Masked autoregressive flow for density estimation,''
\newblock in {\em Advances in NIPS}, 2017, pp. 2338--2347.

\bibitem{shi_aishell-3_2021}
Yao Shi, Hui Bu, Xin Xu, Shaoji Zhang, and Ming Li,
\newblock ``{AISHELL}-3: {A} {Multi}-{Speaker} {Mandarin} {TTS} {Corpus},''
\newblock in {\em Proceedings of Interspeech}, 2021, pp. 2756--2760.

\end{thebibliography}

\end{document}